\documentclass[preprint,pdf]{iucr}              % DO NOT DELETE THIS LINE

     \paperprodcode{a000000}              % Replace with production code if known
     \paperref{xx9999}                    % Replace xx9999 with reference code if known
     \papertype{FA}                       % Indicate type of article
     \journalcode{S}              % Indicate the journal to which submitted
\usepackage{amsmath} %fleqn : align "equations" left
\usepackage{amsfonts}
\usepackage{algorithm}
\usepackage{algpseudocode} % modernized algorithmicx bundle 
\usepackage{setspace} %setstretch
\renewcommand{\algorithmicrequire}{}

\usepackage{amssymb}
\usepackage{graphicx} %for includegraphics scaling
\newcommand{\argmin}[1]{\underset{#1}{\operatorname{argmin}} \; } 

\newcommand{\norm}[1]{\left\| #1 \right\|}
\newcommand{\diff}{\operatorname{d}\!}
\newcommand{\amin}[2]{\underset{#1}{\operatorname{argmin}} \; \left\{ #2 \right\} } 

\begin{document}                  % DO NOT DELETE THIS LINE

     %-------------------------------------------------------------------------
     % The introductory (header) part of the paper
     %-------------------------------------------------------------------------

     % The title of the paper. Use \shorttitle to indicate an abbreviated title
     % for use in running heads (you will need to uncomment it).

\title{A practical local tomography reconstruction algorithm based on known subregion}

%\shorttitle{Short Title}

     % Authors' names and addresses. Use \cauthor for the main (contact) author.
     % Use \author for all other authors. Use \aff for authors' affiliations.
     % Use lower-case letters in square brackets to link authors to their
     % affiliations; if there is only one affiliation address, remove the [a].

\cauthor[a,b]{Pierre}{Paleo}{pierre.paleo@esrf.fr}{}
\author[c]{Michel}{Desvignes}{}%{michel.desvignes@gipsa-lab.grenoble-inp.fr}{}
\cauthor[a]{Alessandro}{Mirone}{mirone@esrf.fr}{}
\aff[a]{ESRF, 71 avenue des Martyrs, 38000 Grenoble \country{France}}
\aff[b]{Universit\'{e}  de Grenoble, Gipsa-Lab, 11 Rue des Math\'{e}matiques, 38400 Saint-Martin-d'H\`{e}res, \country{France}}
\aff[c]{GIPSA-lab - Grenoble Images Parole Signal Automatique - Institut Polytechnique de Grenoble \country{France}}

     % Use \shortauthor to indicate an abbreviated author list for use in
     % running heads (you will need to uncomment it).

%\shortauthor{Soape, Author and Doe}

     % Use \vita if required to give biographical details (for authors of
     % invited review papers only). Uncomment it.

%\vita{Author's biography}

     % Keywords (required for Journal of Synchrotron Radiation only)
     % Use the \keyword macro for each word or phrase, e.g. 
     % \keyword{X-ray diffraction}\keyword{muscle}

%\keyword{keyword}

     % PDB and NDB reference codes for structures referenced in the article and
     % deposited with the Protein Data Bank and Nucleic Acids Database (Acta
     % Crystallographica Section D). Repeat for each separate structure e.g
     % \PDBref[dethiobiotin synthetase]{1byi} \NDBref[d(G$_4$CGC$_4$)]{ad0002}

%\PDBref[optional name]{refcode}
%\NDBref[optional name]{refcode}

\maketitle                        % DO NOT DELETE THIS LINE

\begin{synopsis}
%Supply a synopsis of the paper for inclusion in the Table of Contents.
This paper proposes a method for reducing the cupping effect in local tomography based on a known region.

\end{synopsis}

\begin{abstract}
We propose a new method to reconstruct data acquired in a local tomography setup.
This method uses an initial reconstruction and refines it by correcting the low frequency artifacts known as the cupping effect.
A basis of Gaussian functions is used to correct the initial reconstruction.
The coefficients of this basis are iteratively optimized under the constraint of a known subregion.
Using a coarse basis reduces the degrees of freedom of the problem 
while actually correcting the cupping effect.
Simulations show that the known region constraint yields an unbiased reconstruction, 
in accordance to uniqueness theorems stated in local tomography.
\end{abstract}

     %-------------------------------------------------------------------------
     % The main body of the paper
     %-------------------------------------------------------------------------
     % Now enter the text of the document in multiple \section's, \subsection's
     % and \subsubsection's as required.

\section{Introduction}
In this section, we briefly recall the Region of Interest (ROI) tomography problem and review the related work.
\subsection{Region of Interest Tomography}
\begin{figure}
\begin{center}
\includegraphics[width=0.6\textwidth]{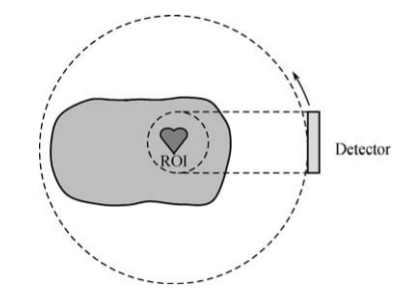}
\end{center}
\caption{
Local tomography setup when the detector covers only a ROI of the object. Image: \cite{book_tutorial}
}
\label{fig:local0}
\end{figure}

Region of Interest (ROI) tomography, also called local tomography, naturally arises when 
imaging objects that are too large for the detector field of view (FOV), as depicted on Figure \ref{fig:local0}.
It notably occurs in medical imaging, where only a small part of a body is imaged. 
Local tomography can also originate from a radiation dose concern in medical imaging.
%\pic{0.6}{images/medicalimrecons/localp31.png}
%{
%Local tomography setup when the detector covers only a ROI of the object. Image: \cite{book_tutorial}\label{fig:local0}
%}

Since the projection data does not cover the entire object, it is said to be \textit{truncated} with respect to a scan that would cover the entire object.
The aim is then to reconstruct the ROI from this ``truncated" data.

However, due to the nature of the tomography acquisition, the acquired data is not sufficient to reconstruct the ROI in general:
for each angle, rays go through the entire object, not only the ROI.
Thus, the data does not only contain information on the ROI, but also contribution from the parts of the object external to the ROI.
For example, on Figure \ref{fig:local0}, the detector gets data from parts of the object located at the left of the ROI.
These contributions from the external parts actually preclude from reconstructing exactly the ROI from the acquired data in general.

The problem of reconstructing the interior of an object from truncated data is referred as the \textit{interior problem}.
It is well known that the interior problem does not have a unique solution in general.
If $P$ denotes the projection operator, $d$ the acquired data
and $x$ a solution of the problem $P(x) = d$, then
$x$ is defined up to a set of \textit{ambiguity functions} $u$ such that $P(x + u) = d$.
An example is given in \cite{local_tomoreconst21} where $u$ is non-zero in the ROI, but $P(u) = 0$ in the detector zone corresponding to the ROI : 
two solutions differing by $u$ would produce the identical interior data.
In \cite{local_meaning}, it is emphasized that the ambiguity is an \textit{infinitely differentiable} function whose variation increases when going outside the ROI.
The non-uniqueness of the solution of the interior problem prevents quantitative analysis of the reconstructed slices.

Methods tackling the ROI tomography problem can mainly be classified in two categories.
The first category methods aim at completing the data by extrapolating the sinogram.
There are often oriented toward easy and practical use, although having no theoretical guarantees.
The second category of methods rely on prior knowledge on the object.
Many theoretical efforts were made on these methods, providing for example uniqueness and stability results.

Other works 
use wavelets to localize the Radon transform \cite{local_wavelet_multires} \cite{local_convolution_backproj}
or focus on the detection of discontinuities, the best known being probably Lambda-tomography \cite{local_lambda_fan}.

\subsection{Sinogram extrapolation methods} 
In a classical tomography acquisition, the whole object is imaged.
If nothing is surrounding the object, the rays are not attenuated by the exterior of the object ; thus 
the sinogram values for each angle go to zero on the left and right parts (after taking the negative logarithm of the normalized intensity). % "zero" after dark-field correction
In a local tomography acquisition, however, the data is ``truncated" with respect to what would have been a whole scan.
%, as illustrated on Figure \ref{fig:localgeom}. %TODO : this has been commented. The image is redundant
%
%\pic{0.3}{images/local_geom.png}
%{
%Illustration of a ROI tomography acquisition. $x_i$ denotes the ROI and $x_e$ denotes the exterior of the ROI.
%For each angle the rays are attenuated by the object on all the detector bins, due to $x_e$ surrounding $x_i$.
%\label{fig:localgeom}
%}
%TODO this was deleted
%\begin{figure}
%\begin{center}
%\includegraphics[width=0.3\textwidth]{images/local_geom.png}
%\end{center}
%\caption{
%Illustration of a ROI tomography acquisition. $x_i$ denotes the ROI and $x_e$ denotes the exterior of the ROI.
%For each angle the rays are attenuated by the object on all the detector bins, due to $x_e$ surrounding $x_i$.
%}
%\label{fig:localgeom}
%\end{figure}
%
The incompleteness of the data induces artifacts on the reconstructed image.
The first obvious artifact is visible as a bright rim on the exterior of the image.
This bright rim is the result of the abrupt transition in the truncated sinogram: the filtration process suffers from a Gibbs phenomenon.
Another artifact is referred as the \textit{cupping effect}: an unwanted background appears in the reconstructed image, which makes further analysis like segmentation 
challenging. 
These two artifacts occur simultaneously, but they have different causes. 
The bright rim comes from the truncation, while the cupping comes from the contribution of the external part.

\begin{figure}
\begin{center}
\includegraphics[scale=0.35]{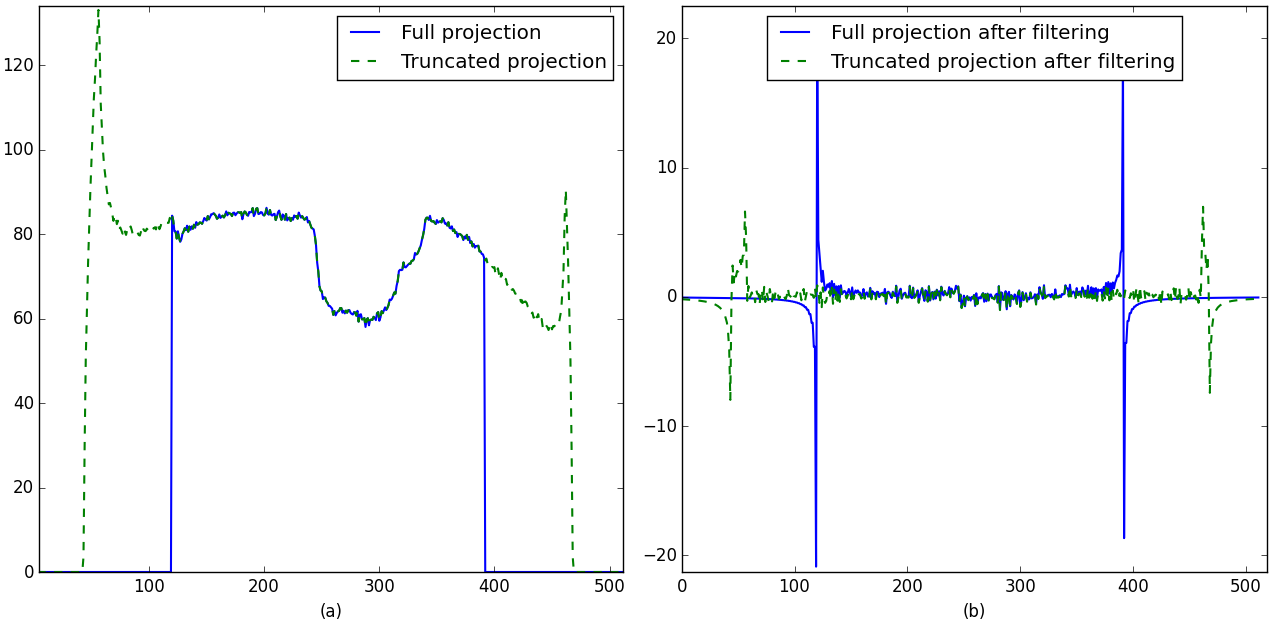}
\end{center}
\caption{
Illustration of the truncation artifacts on a line of the sinogram of the \textit{Shepp-Logan} phantom.
(a): Whole sinogram corresponding to a scan where all the object is imaged (green), and truncated sinogram (blue).
(b): After the ramp-filtering.
}
\label{fig:gibbs}
\end{figure}

\begin{figure}
% --------------------
\begin{minipage}{0.45\textwidth}
\begin{center}
\includegraphics[width=0.7\textwidth]{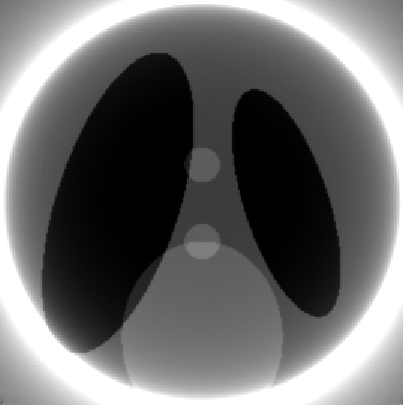}
\\
(a)
\end{center}
\end{minipage}
% --------------------
\begin{minipage}{0.45\textwidth}
\begin{center}
\includegraphics[width=0.9\textwidth]{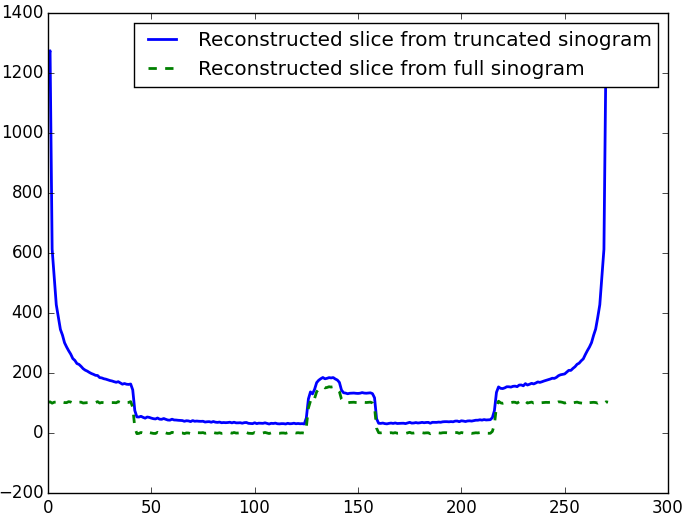}
\\
(b)
\end{center}
\end{minipage}
% --------------------
\caption{
(a): Reconstruction of the truncated sinogram with filtered back-projection. 
The contrast has been modified to visualize the interior of the slice.
(b): Line profile of the reconstruction
}
\label{fig:rim}
\end{figure}

Figure \ref{fig:gibbs} and \ref{fig:rim} illustrates these artifacts. A synthetic slice is projected, and the resulting sinogram is truncated to simulate a ROI tomography setup.
The filtering step enhances the transition between the ROI and the truncated part which is set to zero.
The difference between the filtered whole sinogram and the filtered truncated sinogram also shows the cupping effect, which appears as a 
low-frequency bias.

% ----------------------------------------------------------------------------------------

%\begin{pics}
%\subfig{0.45}{0.7}{images/python/SL_rim.png}{(a)}
%\subfig{0.45}{0.9}{images/python/SL_rim_profile.png}{(b)}
%\caption{
%(a): Reconstruction of the truncated sinogram with filtered back-projection. 
%The contrast has been modified to visualize the interior of the slice.
%(b): Line profile of the reconstruction
%}
%\label{fig:rim}
%\end{pics}

Sinogram extrapolation methods primarily aim at eliminating the bright rim resulting from the truncation
by ensuring a smooth transition between the ROI and the external part.
Besides, efforts have been put into the estimation of the missing data in order to reduce the cupping effect.
These techniques are referred as sinogram extrapolation methods: the external part is estimated from the truncated data with some extrapolating function.

Extrapolating function can be for example constant (the outermost left/right values are replicated), polynomial, $\cos^2$.
In \cite{local_mixed_extrapolations}, a mixture of exponential and quadratic functions are used to estimate the external part, possibly iteratively.
Projection of a circle, for which a closed-form formula is known, can also be used \cite{local_thesis_truncated_projs}.
A common approach is using the values of the left/right part of the sinogram to estimate the external part, that is, replicating the borders values.

In general, sinogram extrapolation methods do not take into account the sinogram theoretical properties.
For example, given an object being nonzero only inside a circle of a given radius,
the sinogram decreases to zero at the left and right boundaries.
Generally speaking, a sinogram of complete measurements satisfies the Helgason-Ludwig consistency conditions \eqref{hl1} \cite{local_thesis_truncated_projs}:
\begin{equation}\label{hl1}
H_n (\theta) = \int_{-\infty}^\infty s^n p(\theta, s) \diff s
\end{equation}
is a homogeneous polynomial of degree $n$ in $\sin \theta$ and $\cos \theta$, for all $n \geq 0$.
An alternative formulation is given by equation \eqref{hl2} :
\begin{equation}\label{hl2}
H_{n, k} (\theta) = \int_0^\pi \int_{-\infty}^\infty s^n e^{j k \theta} p(\theta, s) \diff s \diff \theta \; = \, 0
\end{equation}
for $k > n \geq 0$ and $k - n$ even. 
In \cite{local_ellipse}, \eqref{hl2} is used as a quantitative measure of the sinogram consistency, and is optimized as an objective function.

For many applications, constant extrapolation provides acceptable results \cite{roifbp}, although cupping artifact makes the segmentation challenging.

\subsection{Prior knowledge based interior tomography}\label{priortomo}

It was long believed that ROI tomography cannot be solved exactly,
because of the nature of Radon inversion through FBP: the reconstruction of each voxel 
requires the knowledge of all the (complete) lines passing through this voxel.
However, in the last decade, it has been shown that
multiple nonequivalent reconstruction formulas allow partial reconstruction from partial data in the 2D case \cite{local_tomoreconst21}.
Alternatively to Filtered Back Projection reconstruction, which requires complete data, Virtual Fan Beam (VFB) and Differentiated Back-Projection (DBP)
were developed based on the Hilbert projection equality \cite{local_vfb_vs_dbp}.

Moreover, uniqueness theorems based on analytical continuation of the Hilbert Transform \- were stated and 
progressively refined in \cite{local_1}, \cite{local_2}, \cite{local_2b}, \cite{local_3}, \cite{local_4}, \cite{local_truncated_ht}, \cite{local_apriori_tiny}, \cite{local_7}.
They ensure an exact and stable reconstruction of the ROI given some assumptions.
These assumptions can be of geometric nature, or in the form of a prior knowledge.

% -----------------------------------------------------------------------------------------
\begin{figure}
% --------------------
\begin{minipage}{0.49\textwidth}
\begin{center}
\includegraphics[width=0.9\textwidth]{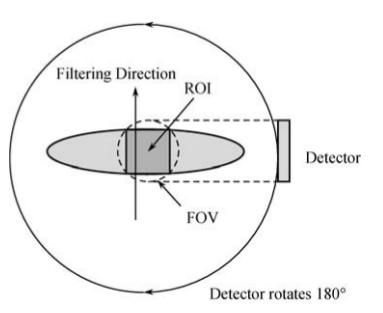}\\
(a)
\end{center}
\end{minipage}%
% --------------------
\begin{minipage}{0.49\textwidth}
\begin{center}
\includegraphics[width=0.9\textwidth]{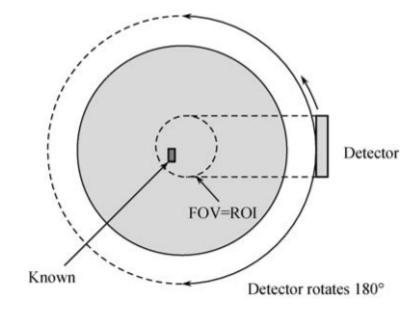}\\
(b)
\end{center}
\end{minipage}
% --------------------
\caption{
(a): Setup where the DBP can reconstruct the ROI. 
As the scanner field of view extends the ROI on both sides, the finite inverse Hilbert Transform can be computed. 
(b): Setup of interior tomography when the FOV does not extend the object. Only the knowledge of a sub-region can provide an exact reconstruction.
Images: \cite{book_tutorial}
}
\label{fig:dbp}
\end{figure}
% ----------------------------------------------------------------------------------------

%\begin{pics}
%\subfig{0.49}{0.9}{images/medicalimrecons/localp32.png}{(a)}
%\subfig{0.49}{0.9}{images/medicalimrecons/localp33b.png}{(b)}
%\caption{
%(a): Setup where the DBP can reconstruct the ROI. 
%As the scanner field of view extends the ROI on both sides, the finite inverse Hilbert Transform can be computed. 
%(b): Setup of interior tomography when the FOV does not extend the object. Only the knowledge of a sub-region can provide an exact reconstruction.
%Images: \cite{book_tutorial}
%\label{fig:dbp}
%}
%\end{pics}

Geometry-based prior knowledge is related to the acquisition geometry. For example, in DBP based reconstruction,
a point can be reconstructed if it lies on a line segment extending outside the object on both sides, and all lines crossing the segment are measured
\cite{local_tomoreconst21}, as shown in Figure \ref{fig:dbp} (a). 
Similar results were obtained under less restrictive assumptions, for example the field of view extending the ROI on only one side \cite{local_truncated_ht}.

These geometry-based methods do not work, however, when the FOV does not extend the object (Figure \ref{fig:dbp} (b)).
In this case, it has been shown \cite{local_apriori}
that a prior knowledge on the function inside the ROI enables exact and stable reconstruction of the ROI.
This knowledge can be in the form of the function values inside a sub-region of the ROI \cite{local_apriori_tiny}
or can be about the properties of the function to reconstruct, for example sparsity in some domain.

This latest kind of knowledge has led to compressive sensing based ROI tomography.
In \cite{local_highordertv}, \cite{local_cs_based}, \cite{local_generalized_tv}, Total Variation method is used to reconstruct the ROI.
In \cite{local_wavelet_sparsity_besov}, the function is assumed to be sparse in the wavelet domain, and a multi-resolution scheme reduces
the number of unknown by keeping only fine-scale wavelet coefficients inside the ROI.
In \cite{local_wavelets_weighted_sparsity}, it is shown that piecewise constant functions are determined \textit{everywhere} by their ROI data, the underlying hypothesis being 
formulated as sparsity in the Haar domain.

\section{Low frequencies artifacts correction with Gaussian blobs}

Sinogram extrapolation usually copes well with the correction of discontinuities in the truncated sinogram,
but does not correct the cupping effect in general.
This cupping effect appears as a low frequency bias in the reconstructed image. 
Sinogram extrapolation and other background correction techniques do not give guarantees that the low frequency bias will actually be removed without distorting  the reconstruction.

In this section, we describe a new method using prior knowledge on a subregion of the reconstructed volume to eliminate the low frequency cupping bias.
The starting point of this method is an initial reconstruction, hereby denoted $x_0$, which can be obtained for example with the padded FBP method. 
This initial reconstruction is then refined with an additive correction term.
This correction term uses the known sub-region as a constraint 
which should be sufficient, according to uniqueness theorems stated in the references given in \ref{priortomo}, 
to accurately reconstruct the region of interest.

As $x_0$ bears the high frequencies features, the correction term 
is expressed as a linear combination of Gaussian functions to counterbalance the low frequency artifacts.
The coefficients are optimized subject to the knowledge of the subregion, hereby denoted $\Omega$.
To constrain all the Gaussian coefficients by the knowledge of the image values in $\Omega$,
a reduced set of coefficients is firstly computed inside $\Omega$. 
Then, the Gaussian coefficients are iteratively optimized to fit the reconstruction error of the whole image,
using the coefficients computed inside $\Omega$ as a constraint.

Let $u_0$ denote the ``true" object values in the known region $\Omega$. The proposed method can be summarized as follows:
\begin{itemize}
\item The reconstruction error in $\Omega$, denoted $e_{|\Omega} = \left( (x_0)_{|\Omega} - u_0 \right)$, is expressed as a linear combination of two dimensional Gaussians. 
The resulting Gaussian coefficients are denoted $g_0$.
\item The error in the whole image is iteratively fitted with Gaussians coefficients $g$, subject to $g_{|\Omega} = g_0$, to build a consistent reconstruction error in the whole image.
\end{itemize}
Details of each step are described in the following parts.

\subsection{Capturing the low frequencies of the error in the known zone}\label{part1}
The key assumption of this method is that $\Omega$ is large enough to bear sufficient information on the low frequencies artifacts (cupping effect) of a classical reconstruction.
The reconstruction error in $\Omega$ is approximated as a linear combination of Gaussian functions. 
This function is chosen for computational convenience, more details are given in \ref{comp}.

Equation \eqref{g1d} gives the expression of the approximation.% in the one dimensional case, the extension to two dimensions being straightforward.
The error estimation $\widehat{e_{|\Omega}}$ is a linear combination of translated Gaussians of weights $c_{i,j}$, with a spacing $s$.

\begin{equation}\label{g1d}
\begin{aligned}
\widehat{e_{|\Omega}} &= \sum_{i, j} c_{i,j} \psi_\sigma (x - i \cdot s, y - j \cdot s)
\\
\psi_\sigma (x, y) &= \frac{1}{\sigma \sqrt{2\pi}} \exp \left( -\frac{x^2 + y^2}{2\sigma^2} \right)
\end{aligned}
\end{equation}

For simplicity, all the Gaussians have the same standard deviation $\sigma$ and all have the same spacing $s$ between them. 
Their location on the image grid is also fixed, so that the fit turns into a linear inverse problem :
\begin{equation}\label{opt0}
g_0 = \amin{g}{\frac{1}{2}\norm{G g - \left( (x_0)_{|\Omega} - u_0 \right)}_2^2}
\end{equation}
where $G$ is the operator taking as an input the coefficients $c_i$, stacked in a vector $g$ ; and producing an image tiled with Gaussians (here in region $\Omega$).
The norm $\norm{\cdot}_2^2$ is the squared Frobenius norm, that is, the sum of the squares of all components.
Choosing the same standard deviation $\sigma$ and the same spacing $s$ for all Gaussians enables to implement $G$ as a convolution. 
More precisely, given an image being zero everywhere, coefficients $c_i$ are placed every $s$ pixels in $\Omega$
and this image is convolved with two dimensional $\psi_\sigma$.

The reconstruction error in $\Omega$ is thus estimated in the least squares sense: $g_0$ is the vector of Gaussian coefficients
giving the best estimation $\widehat{e_{|\Omega}}$ of the reconstruction error in the L2 sense.
This vector $g_0$ will be used in the second part of the algorithm.

\subsection{Correcting the reconstruction error in the image}\label{part2}
\subsubsection{Overview of the method}
Outside the known region $\Omega$, the reconstruction error is not known.
Like some other methods described in \ref{priortomo}, this algorithm aims at using the known region information
to accurately reconstruct the whole ROI.
However, this approach focuses on \textit{correcting} an initial reconstruction: the reconstruction error in $\Omega$
is fitted by as a linear combination of Gaussians, then the whole image is corrected in a coarse Gaussian basis
whose coefficients are constrained in the known subregion.

The Filtered Backprojection (FBP) with sinogram extrapolation is widely used in local tomography
because it is both simple and gives satisfactory results in general \cite{roifbp}.
Theoretical investigations found that FBP provides a reconstructed function bearing the same discontinuities as the reference function
\cite{local_fbp_bilgot} %, as other methods like pseudo-local tomography and $\Lambda$-tomography.
, although the cupping effect can make the segmentation challenging.
In this method, FBP with padding is used to obtain an initial estimate of the reconstruction ; 
the aim is to correct the local tomography artifacts on this image using the prior knowledge.
Equation \eqref{lowfreqs0} gives the expression of the estimate $x$ 
where $x_0$ is the initial reconstruction, $G$ is the operator described in \ref{part1}
and $\hat{g}$ is a linear combination of Gaussian functions aiming at counterbalancing the low frequencies artifacts.
\begin{equation}\label{lowfreqs0}
x = \tilde{x_0} + G \hat{g}
\end{equation}
The vector $\hat{g}$ is found by minimizing an objective function which is built as follows.
A new image $x = \tilde{x_0} + G g$, containing the initial reconstruction, is created.
The image $\tilde{x_0}$ is an extension of the initial reconstruction $x_0$.
This new image is projected with a projector $P$ adapted to the bigger size.
To compare with the acquired sinogram $d$, the computed sinogram $P (\tilde{x_0} + G g)$ is truncated by a cropping operator $C$.
The data fidelity is then given by Equation \eqref{fid1}.
\begin{equation}\label{fid1}
\frac{1}{2} \norm{
C P ( \tilde{x_0} + G g ) - d
}_2^2
\end{equation}

We emphasize that this approach differs from the full estimation of the ROI based on a subregion.
The variables $g$ are in a coarse basis while $\tilde{x_0}$ is fixed, which is notably reducing the degrees of freedom of the problem.
The operation $G g$ has two goals: a coarse estimation of the exterior (outside the $x_0$ support)
and a correction of the low frequencies error inside the $x_0$ support.

As the minimization is on $g$, the initial estimate of the ROI $x_0$ is constant, and the data fidelity term \eqref{fid1} can be rewritten as in Equation \eqref{fid2}
\begin{equation}\label{fid2}
\frac{1}{2}
\norm{
C P G g - d_e
}_2^2
\end{equation}

where $d_e = d - \tilde{P} x_0$ is the difference between the acquired sinogram $d$ and the projection of the initial reconstruction $x_0$,
and $\tilde{P}$ is the projector adapted to the size of $x_0$.
The optimization problem is given by Equation \eqref{opt1}.
\begin{equation}\label{opt1}
\hat{g} = \amin{g}{
\frac{1}{2}
\norm{
C P G g - d_e
}_2^2
\quad \text{subject to } \quad g_{|\Omega_g} = g_0
}
\end{equation}

$g_0$ is the vector of Gaussian coefficients found in \eqref{opt0}, such that $G g_0$ approximates the error in the known zone.
The set $\Omega_g$ denotes the subset of the Gaussian basis corresponding to $\Omega$ in the pixel basis: 
if a coefficient $c_i$ of $g$ lies in $\Omega_g$ in the Gaussian basis, then $(G g)_i$ lies in $\Omega$ in the pixel basis.
Equation \eqref{opt1} boils down to finding coefficients $g$ 
minimizing the reconstruction error in the whole image,
under the constraint that $G g$ should give the (known) reconstruction error in $\Omega$.

This local constraint is propagated in all the variables by the projection operator involved in the optimization process.
Uniqueness theorems mentioned in \ref{priortomo} state that the knowledge of a subregion of the ROI
is sufficient to yield an exact reconstruction.
However, when using a pixel basis without space constraints, the number of degrees of freedom might be too high ; 
leading to a slow convergence.
Using a coarse basis for correcting the low frequencies reduces this number of degrees of freedom.

\subsubsection{Details on the involved operations}\label{opdetails}
In this part, more details are given on the different steps of the algorithm.
We start by computing a padded FBP reconstruction $x_0$ which gives an initial estimate of the ROI of size $(N, N)$.
This image is extended to a bigger image $\tilde{x_0}$ of size $(N_2, N_2)$ where $N_2 > N$, and $x_0$ is placed in the center of the image.

At each iteration $k$, the image $G g_k$, where $g_k$ is the Gaussian coefficients vector at iteration $k$, is computed.
The operator $G$ consists in placing the coefficients $g$ on a regular grid and convolving with the Gaussian kernel 
$(x, y) \mapsto \frac{1}{\sigma\sqrt{2\pi}}\exp \left( -\frac{x^2+y^2}{2\sigma^2}\right)$.
Thus, the operator $G$ can be written $G = C_\sigma U_s$
where $C_\sigma$ is the convolution by the aforementioned Gaussian kernel of standard deviation $\sigma$,
and $U_s$ is an operator upsampling an image by a factor of $s$. % (hereby denoted $s$-upsampling).
As both are linear operators, $G$ is a linear operator and $G^T = U_s^T C_\sigma^T$ where $U_s^T$ is the $s$-downsampling operator 
and $C_\sigma^T$ is a convolution by the matched Gaussian kernel, which is the same kernel due to symmetry.
In our implementation, the Gaussian kernel has a size of $\lfloor 8\sigma + 1\rfloor$, i.e the Gaussian is truncated at $4 \sigma$. %in order to have kernel both accurate and relatively small.
The resulting image is given by Equation \eqref{discreteconv}
\begin{equation}
\begin{aligned}
x(i_0, j_0) &= (G g)(i_0, j_0) = (C_\sigma U_s g)(i_0, j_0) = (C_\sigma z)(i_0, j_0) \\
&= \sum_{i, j} z(i, j) \psi_\sigma (i_0 - i, j_0 - j)
\end{aligned}
\label{discreteconv}
\end{equation}
where $\psi_\sigma$ is the discrete Gaussian kernel, 
$z$ is the image containing the Gaussian coefficients $g$ placed on the grid of size $(N_2, N_2)$ after upsampling, that is,
zeros almost everywhere except coefficients every $s$ pixel. The summation in Equation \eqref{discreteconv} is done on the convolution kernel support.
If $s < 4 \sigma$, the Gaussian functions supports can overlap once placed on the grid. 
In practice, these Gaussians should overlap to appropriately fit constant regions : 
for $s$ close to $\sigma$, the Gaussians almost yield a partition of unity \cite{gaussians_partition_unity}.

% Subsampling matrix:
%\begin{equation*}
%\begin{bmatrix}
%1 & 0 & 0 & \ldots & 0 \\
%\multicolumn{5}{c}{0_{1,s}}  \\
%0 & 1 & 0 & \ldots & 0 \\
%\multicolumn{5}{c}{0_{1,s}}  \\
%0 & 0 & 1 & \ldots & 0 \\
%\multicolumn{5}{c}{\ldots}  \\
%0 & 0 & 0 & \ldots & 1
%\end{bmatrix}
%\end{equation*}

\begin{figure}
\begin{center}
\includegraphics[scale=0.3]{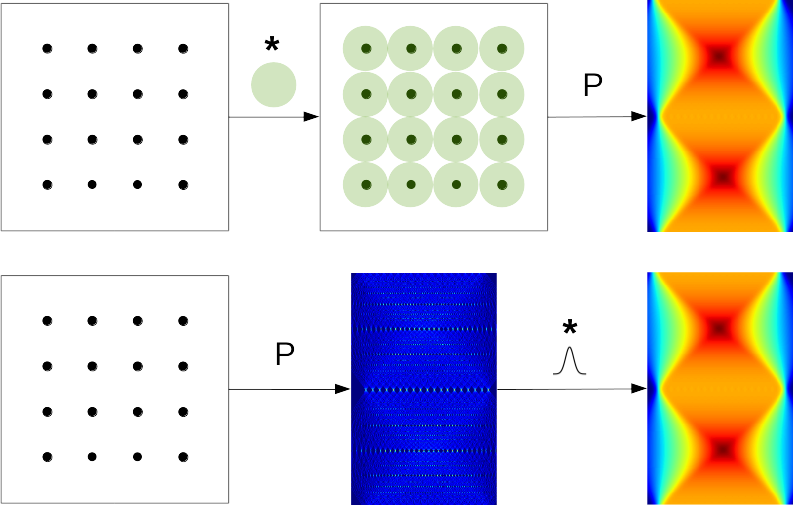}
\end{center}
\caption{
Description of the operator $PG$ (first line).
A grid of points is created ; one coefficient is assigned to each point on the grid.
The result is convolved by a two dimensional Gaussian function (depicted as green circles), and projected to obtain a sinogram.
An equivalent approach (second line) consists in first projecting the ``point coefficients" with an appropriate projector, 
and convolving each line of the sinogram by a one dimensional Gaussian function.
}
\label{fig:conv}
\end{figure}

Once the coefficients are placed on a grid and convolved by the 2D Gaussian function, the image $x$ is projected.
The projection operator adapted to the new geometry (the bigger image $G g_k$) is denoted $P$. This is a standard Radon transform.
This process is illustrated in Figure \ref{fig:conv}.

As the new image $x$ is bigger than $x_0$, the sinogram $P x$ and the acquired data $d$ cannot be directly compared.
The computed sinogram $P x$ is thus cropped to the region corresponding to the ROI.
The cropping operator is denoted by $C$ ; it is also a linear operator whose transpose consists in extending the sinogram by inserting zeros on both sides.

The resulting sinogram aims at fitting the error between the acquired sinogram $d$ and the (cropped) projection of the object.
As the object is unknown except inside $\Omega$, the reconstruction error is only known in $\Omega$.
The Gaussian coefficients $g$ are constrained by those found by fitting the error inside $\Omega$ in \ref{part1}.
The projection operator involved in the process propagates the constraint to all the other coefficients.

% -----------------------------------------------------------------------------------------

Coefficients $\hat{g}$ from Equation \eqref{opt1} are computed with an iterative solver. 
As this objective function is quadratic, efficient minimization algorithms like conjugate gradient can be used.
The final image is obtained with $\hat{x} = x_0 + G \hat{g}$ and is cropped to the region of interest.

\subsubsection{Computational aspects}\label{comp}
Using Gaussians as functions to iteratively express the error has several computational advantages.
Gradient-based algorithms for solving \eqref{opt1} %
involve the computation of the forward operator $P G$ and its adjoint $G^T P^T$.
They are usually the computationally expensive steps of iterative solvers. 
In this case, these operators can be computed in an efficient way.

The Gaussian kernel has an interesting property: it is the only (nonzero) one to be both rotationally invariant and separable \cite{gaussians}.
In our case, the convolution by a Gaussian followed by a projection (forward Radon transform)
is equivalent to projecting first and convolving each line of the sinogram by the corresponding one dimensional Gaussian, as illustrated on Figure \ref{fig:conv}.

The first advantage is of theoretical nature. In many implementations, the projector and backprojector pair are usually not adjoint of each other for performances reasons.
Although giving satisfying results in most practical applications, this raises theoretical issues on convergence of algorithms
using iteratively forward and backward operators \cite{unmatched}. 
By using a point-projector and a point-backprojector implementation, the pair can be exactly adjoint, besides giving more accurate results.

The second advantage is on the computational side.
As the operator $P G$ consists in projecting 2D Gaussians disposed on the image,
it is equivalent to placing one-pixel coefficients (Dirac functions in the continuous case) in the image on a grid denoted $\mathcal{I}$,
projecting the image (with a point-projector) and convolving the sinogram by a one-dimensional Gaussian kernel.
The same goes for the adjoint operator $G^T P^T$ consisting in retrieving the Gaussians coefficients from a sinogram.
The standard way to compute this operator would be backprojecting the sinogram,
convolving by the two dimensional Gaussian (which is its own matched filter due to symmetry), and sampling the image on the grid $\mathcal{I}$ to get the coefficients.
Here, the convolution can be first performed in one dimension along the sinogram lines.
The sinogram is sampled at locations corresponding to points $\mathcal{I}$ in the image domain.
The resulting sampled sinogram is then backprojected with a point-backprojector.

\subsection{Pseudocode of the proposed method}
In this section, the different steps of the proposed method are summarized in two algorithms.
The first performs the fitting of the error in the known zone as described in section \ref{part1}, 
the second builds the resulting image as described in section \ref{part2}.

A complete implementation of the proposed method is available at \cite{gitlocaltomo}.
It contains comments on the different steps and can be tuned for various setups.
This implementation relies on the ASTRA toolbox \cite{astragpu} \cite{astra}, the point-projector scheme described in \ref{comp} is not implemented for readability ; 
but this approach would be more suited to a production reconstruction algorithm where performances are an issue.

% Leave a blank line before algorithm !
\begin{algorithm}[H]\setstretch{1.35}
Algorithm 1. Known zone fitting.\\ % Emulate caption which is not working with this template for some reason
\caption{Known zone fitting}\label{alg:algo1}
\algorithmicrequire
$d$: acquired sinogram \\
$\Omega$: location and size of the known zone \\
$u_0$: known values in the zone $\Omega$
\begin{algorithmic}[1]
\Procedure {fitknown}{$d$, $\Omega$, $u_0$}
\State $x_0 = \operatorname{padded}\_\operatorname{FBP}(d)$ \Comment{Padded FBP of $d$}
\State $e_{|\Omega} = (x_0)_{|\Omega} - u_0$ \Comment{Error in the known zone}
\State $g_0 = \amin{g}{\norm{G g - e_{|\Omega}}_2^2}$ \Comment{Fit the error with Gaussians}
\State $\tilde{x_0} = \operatorname{extend}(x_0)$ \Comment{Extend to a bigger image}
\State \Return $\tilde{x_0},\, g_0$
\EndProcedure
\Statex
\end{algorithmic}
\end{algorithm}

The location of the known zone $\Omega$ can be simply implemented as a tuple of pixels $(i_0, j_0)$ and a radius $r$ for a circular zone.

% Leave a blank line before algorithm !
\begin{algorithm}[H]\setstretch{1.35}
Algorithm 2. Error correction \\ % Emulate caption which is not working with this template for some reason
\caption{Error correction}\label{alg:algo2}
\algorithmicrequire
$d$: acquired sinogram \\
$\Omega$: location and size of the known zone \\
$u_0$: known values in the zone $\Omega$ \\
$N_2$: size of the extended image \\
$\sigma$: standard deviation of the Gaussian functions \\
$s$: grid spacing
\begin{algorithmic}[1]
\Procedure {localtomo}{$d$, $\Omega$, $u_0$, $N_2$, $\sigma$, $s$}
\State $\tilde{x_0},\, g_0 = \operatorname{FITKNOWN}(d, \Omega, u_0)$ \Comment{Compute $\tilde{x_0}$ and $g_0$ with Algorithm 1}% \ref{alg:algo1}}
%\State $x_0 = \operatorname{zeros}(N_2, N_2)$, place $\tilde{x_0}$ in the center of $x_0$ \Comment{Compute the extended image}
\State $d_e = C P \tilde{x_0} - d$ \Comment{Difference between the cropped projection of $x_0$ and $d$}
\State $
\hat{g} = \amin{g}{
\frac{1}{2}
\norm{
C P G g - d_e
}_2^2
\quad \text{s.t.} \quad g_{|\Omega_g} = g_0
}
$ \Comment{Operators are described in \ref{opdetails}}
\State $\hat{x} = \tilde{x_0} + G \hat{g}$
\State \Return $\hat{x}$
\EndProcedure
\Statex
\end{algorithmic}
\end{algorithm}
In practice, the final image $\hat{x}$ is cropped to the region of interest. 
In algorithm 2, %\ref{alg:algo2}, 
the optimization (line 5) can be done with a gradient algorithm, as differentiating the quadratic error term
requires only the operators and their adjoints.

\section{Results and discussion}
In this section, results and discussions on three test cases are presented.
Synthetic sinograms are generated by projecting an object and truncating the sinogram to the radius of a given region of interest in the image.

The following notations are used: 
$\sigma$ is the standard deviation of the Gaussians of the basis, $s$ is the grid spacing, $N_2$ is the size (width or height in pixels) of the extended image
and $R$ is the radius (in pixels) of the known region.
In practice, the size of the ``original image" (which corresponds to the size of an image that would contain the whole object in practice)
is unknown, hence $N_2$ is always chosen different from the width of the original test image. % This avoids the "inverse crime"

In all cases, the input image is projected with a projector covering the entire object.
The resulting sinogram is then truncated to the radius of the region of interest. The truncated sinogram is the input of the methods.
The proposed method is compared to the padded FBP. 
As the padded FBP is used as an initial reconstruction by the proposed method, 
the benchmark is mainly about checking that the cupping effect is actually removed, and that the correction does not induce distortion to the final image.

% -----------------------------------------------------------
% -------------------- First test----------------------------
% -----------------------------------------------------------

\begin{figure}
% --------------------
\begin{minipage}{0.45\textwidth}
\begin{center}
\includegraphics[width=0.7\textwidth]{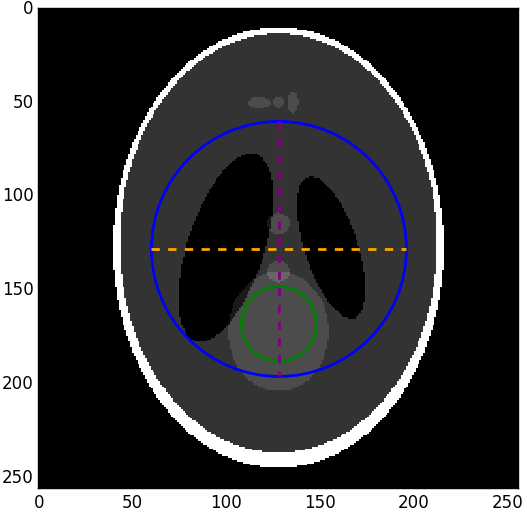}\\
(a)
\end{center}
\end{minipage}%
% --------------------
\begin{minipage}{0.45\textwidth}
\begin{center}
\includegraphics[width=0.7\textwidth]{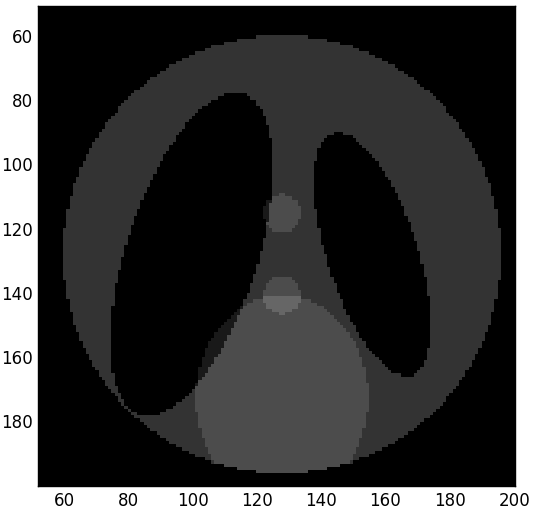}\\
(b)
\end{center}
\end{minipage}
% --------------------
\caption{
(a): Shepp-Logan phantom of size $256 \times 256$.
The outer circle is the region of interest, the inner circle is the known subregion.
The dashed lines indicate the profiles which are to be plotted in the reconstructed slice.
(b): View of the region of interest.
}
\label{fig:sl256}
\end{figure}

The first test involves the standard Shepp-Logan phantom (Figure \ref{fig:sl256}), $256 \times 256$ pixels.
The region of interest is embedded inside the ``absorbing outer material" (ellipse with the highest gray values) to simulate a local tomography acquisition.
For an easier interpretation of the line profiles in the final reconstructed images, the values of the standard phantom are multiplied by $250$ so that all the values are between $0$ and $250$. The width of the extended image is $N_2 = 260$.

% -----------------------------------------------------------------------------------------
% ----------------------------------------------------------------------------------------

%\begin{pics}
%\subfig{0.45}{0.7}{images/results/SL256_3.png}{(a)}
%\subfig{0.45}{0.7}{images/results/volint_SL_4.png}{(b)}
%\caption{
%(a)
%Shepp-Logan phantom of size $256 \times 256$.
%The outer circle is the region of interest, the inner circle is the known subregion.
%The dashed lines indicate the profiles which are to be plotted in the reconstructed slice.
%(b) View of the region of interest.
%}
%\label{fig:sl256}
%\end{pics}

\begin{figure}
% --------------------
\begin{minipage}{0.4\textwidth}
\begin{center}
\includegraphics[width=0.9\textwidth]{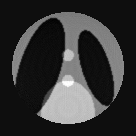}\\
(a)
\end{center}
\end{minipage}%
% --------------------
\begin{minipage}{0.4\textwidth}
\begin{center}
\includegraphics[width=0.9\textwidth]{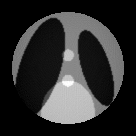}\\
(b)
\end{center}
\end{minipage}
% --------------------
\caption{
Results of reconstructions. (a) Proposed (b) padded FBP
}
\label{fig:sl1}
\end{figure}

Figure \ref{fig:sl1} shows the result of the reconstruction with padded FBP and with the proposed method. 
The Gaussian coefficients were computed with $\sigma = 4$ on a grid of spacing $s = 6$. The known region radius is $R = 20$ pixels, and the extended image width is $N_2 = 260$ pixels.

\begin{figure}
% --------------------
\begin{minipage}{0.45\textwidth}
\begin{center}
\includegraphics[width=1.0\textwidth]{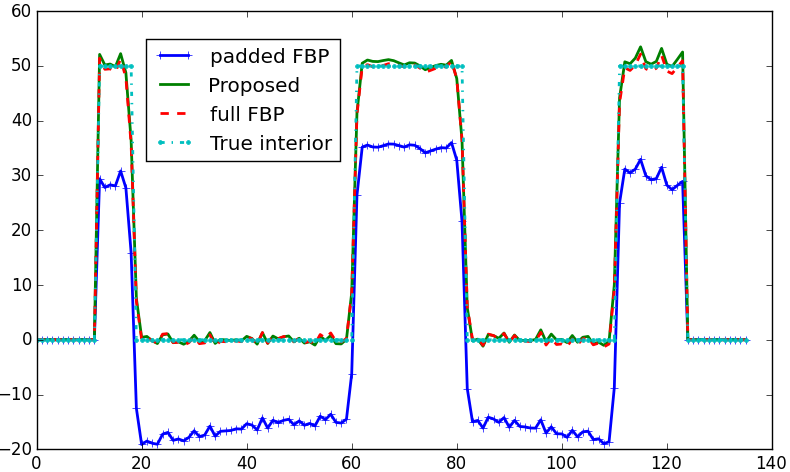}\\
(a)
\end{center}
\end{minipage}%
% --------------------
\begin{minipage}{0.45\textwidth}
\begin{center}
\includegraphics[width=0.9\textwidth]{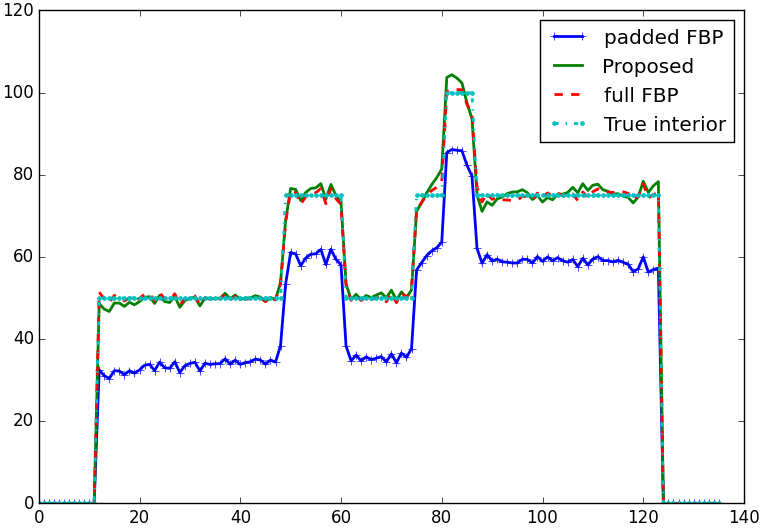}\\
(b)
\end{center}
\end{minipage}
% --------------------
\caption{
Line profiles of reconstructions.
(a): Middle line. (b): Middle column
}
\label{fig:line1}
\end{figure}

By visual inspection, this method do not induce new artifacts in the reconstruction.
Figure \ref{fig:line1} shows a line profile of this reconstruction. The cupping effect is visible for the padded FBP,
and it has been removed with the proposed method.
More importantly, the average reconstructed values are distributed around the true interior values. 
This provides an illustration of the uniqueness theorem: knowing the values of a subregion of the ROI
enables to exactly reconstruct (up to numerical errors) the ROI.
The reconstruction with the proposed method bears the same high frequencies as the FBP with full data,
which is a good indication that this method do not induce new artifacts.
The fact that the reconstruction has the same mean values than the true interior
could enable quantitative analysis of the reconstructed volume, which is not easily achievable in local tomography.

% -----------------------------------------------------------------------------------------
% ----------------------------------------------------------------------------------------

%\begin{pics}
%\subfig{0.4}{0.9}{images/results/SL_proposed1.png}{(a)}
%\subfig{0.4}{0.9}{images/results/SL_FBP1.png}{(b)}
%\caption{
%Results of reconstructions. (a) Proposed (b) FBP
%}
%\label{fig:sl1}
%\end{pics}

% -----------------------------------------------------------------------------------------
% ----------------------------------------------------------------------------------------

%\begin{pics}
%\subfig{0.45}{1.0}{images/results/SL_sig4s6midlinepr.png}{(a)}
%\subfig{0.45}{0.9}{images/results/SL_sig4s6midcolpr.png}{(b)}
%\caption{
%Line profiles of reconstructions.
%(a): Middle line. (b): Middle column
%%The proposed method bears the same discontinuities as the interior while having an almost null mean bias, that is, no cupping effect.
%}
%\label{fig:line1}
%\end{pics}

\begin{figure}
% --------------------
\begin{minipage}{0.45\textwidth}
\begin{center}
\includegraphics[width=1.0\textwidth]{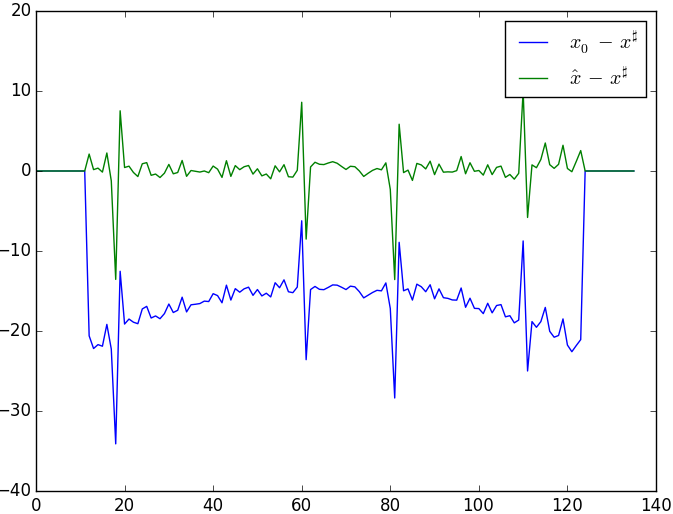}\\
(a)
\end{center}
\end{minipage}%
% --------------------
\begin{minipage}{0.45\textwidth}
\begin{center}
\includegraphics[width=0.9\textwidth]{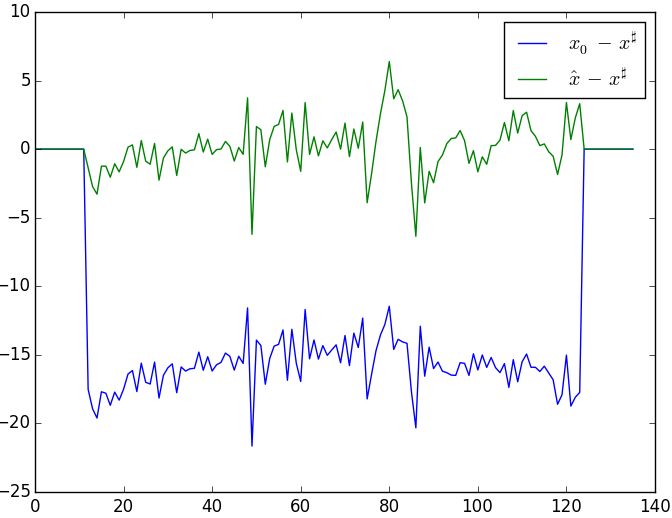}\\
(b)
\end{center}
\end{minipage}
% --------------------
\caption{
Difference between the reconstruction and the true volume $x^\sharp$ 
(a): along the middle line,
(b): along the middle column
}
\label{fig:line2}
\end{figure}

Figure \ref{fig:line2} shows the difference between the reconstructions and the interior values (denoted $x^\sharp$).
As expected, the cupping effect is visible for padded FBP, while being almost entirely suppressed in the reconstruction with proposed method.

% -----------------------------------------------------------------------------------------
% ----------------------------------------------------------------------------------------

%\begin{pics}
%\subfig{0.45}{1.0}{images/results/SL_sig4s6midline.png}{(a)}
%\subfig{0.45}{0.9}{images/results/SL_sig4s6midcol.png}{(b)}
%\caption{
%Difference between the reconstruction and the true volume $x^\sharp$ 
%(a): along the middle line,
%(b): along the middle column
%}
%\label{fig:line2}
%\end{pics}

% -----------------------------------------------------------
% -------------------- Second test---------------------------
% -----------------------------------------------------------

\begin{figure}
% --------------------
\begin{minipage}{0.49\textwidth}
\begin{center}
\includegraphics[width=0.9\textwidth]{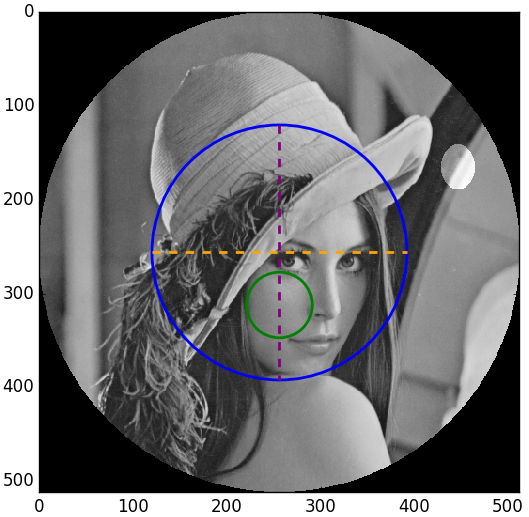}\\
(a)
\end{center}
\end{minipage}%
% --------------------
\begin{minipage}{0.49\textwidth}
\begin{center}
\includegraphics[width=0.9\textwidth]{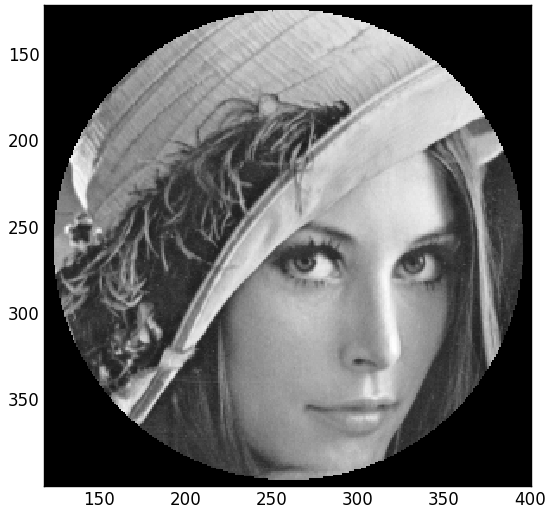}\\
(b)
\end{center}
\end{minipage}
% --------------------
\caption{
(a) Phantom ``Lena". An ellipse with high gray values has been added to accentuate the local tomography setup.
The outer circle is the ROI, and the inner circle is the known region.
The dashed lines indicate the profiles which are to be plotted in the reconstructed slice.
(b) View of the region of interest.
}
\label{fig:lena512_ph}
\end{figure}

The second test involves the test image ``Lena", $512 \times 512$ pixels, bearing both smooth regions and high frequencies textures.
Figure \ref{fig:lena512_ph} shows the test setup. 
The known region has be chosen as slowly varying as possible,
as in real acquisitions the known region is likely to be air or coarse features. The width of the extended image is $N_2 = 520$.

% -----------------------------------------------------------------------------------------
% ----------------------------------------------------------------------------------------

%\begin{pics}
%\subfig{0.49}{0.9}{images/results/Lena512_ph_2.png}{(a)}
%\subfig{0.49}{0.9}{images/results/volint_lena_2}{(b)}
%\caption{
%(a) Phantom ``Lena". An ellipse with high gray values has been added to accentuate the local tomography setup.
%The outer circle is the ROI, and the inner circle is the known region.
%The dashed lines indicate the profiles which are to be plotted in the reconstructed slice.
%(b) View of the region of interest.
%}
%\label{fig:lena512_ph}
%\end{pics}

\begin{figure}
% --------------------
\begin{minipage}{0.3\textwidth}
\begin{center}
\includegraphics[width=0.9\textwidth]{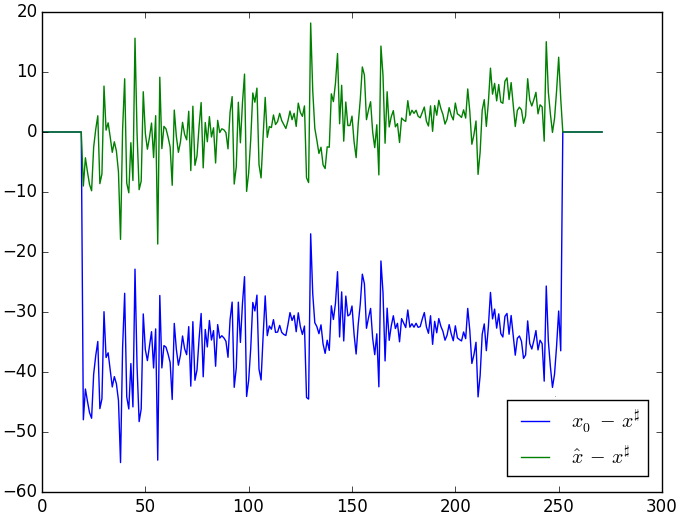}\\
(a)
\end{center}
\end{minipage}%
% --------------------
\begin{minipage}{0.3\textwidth}
\begin{center}
\includegraphics[width=0.9\textwidth]{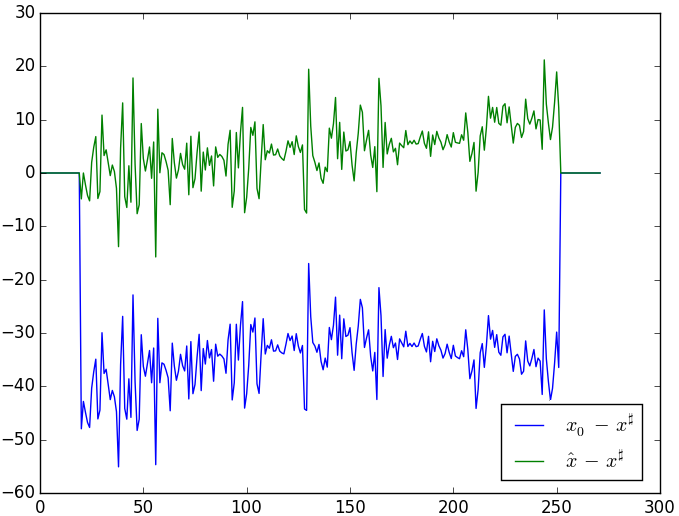}\\
(b)
\end{center}
\end{minipage}
% --------------------
\begin{minipage}{0.3\textwidth}
\begin{center}
\includegraphics[width=0.9\textwidth]{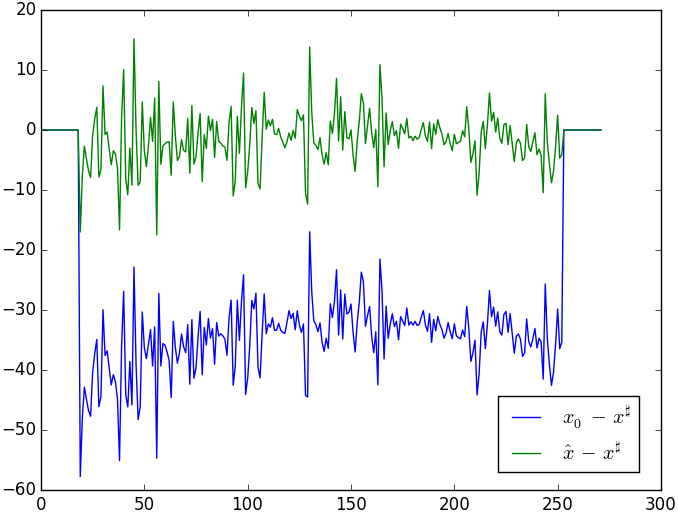}\\
(c)
\end{center}
\end{minipage}
% --------------------
\textbf{\\}
\begin{minipage}{0.3\textwidth}
\begin{center}
\includegraphics[width=0.9\textwidth]{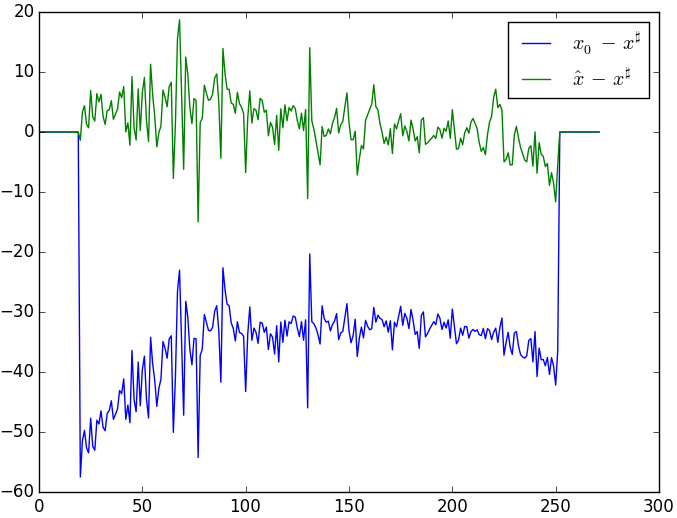}\\
(d)
\end{center}
\end{minipage}%
% --------------------
\begin{minipage}{0.3\textwidth}
\begin{center}
\includegraphics[width=0.9\textwidth]{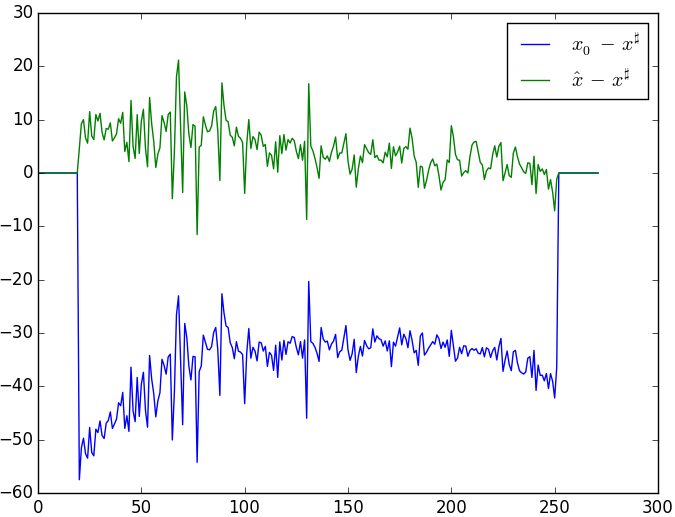}\\
(e)
\end{center}
\end{minipage}
% --------------------
\begin{minipage}{0.3\textwidth}
\begin{center}
\includegraphics[width=0.9\textwidth]{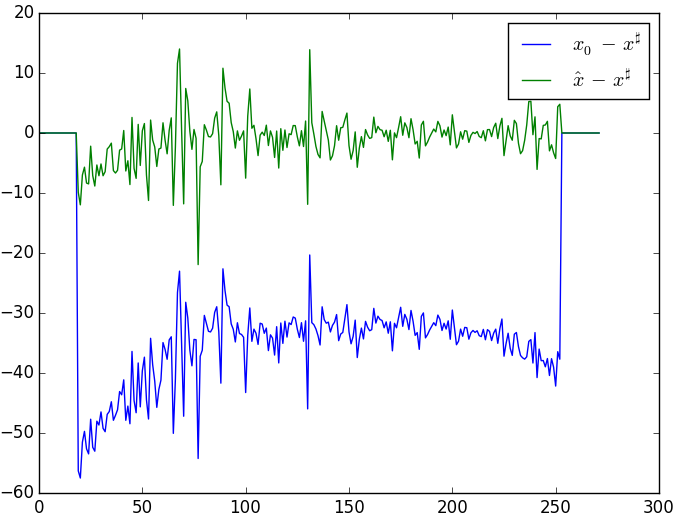}\\
(f)
\end{center}
\end{minipage}
% --------------------
\caption{
Profiles of difference between the reconstruction and the true interior for the \textit{Lena} image.
$x_0$, $\hat{x}$ and $x^\sharp$ are the padded FBP, the proposed reconstruction and the true interior, respectively.
In blue: difference between the padded FBP and the true interior. In green: difference between the reconstruction with the proposed method with $\sigma = s = 3$ and the true interior.
First row: profiles of the middle line of the image for (a) $R = 35$, (b) $R = 15$, (c) $R = 50$.
Second row: profiles of the middle column for (d) $R = 35$, (e) $R = 15$, (f) $R = 50$
}
\label{fig:lenadiff}
\end{figure}

Figure \ref{fig:lenadiff} shows the difference between the true interior and the reconstruction with the proposed method, with varying values of the radius $R$ of the known region.
The parameters $\sigma = s = 3$ has been used for these reconstructions.
As it can be expected, the cupping effect removal is better when the known region is wide.

% -----------------------------------------------------------------------------------------
% ----------------------------------------------------------------------------------------

%\begin{pics}
%\subfig{0.3}{0.9}{images/results/Lena512_midline6.png}{(a)}
%\subfig{0.3}{0.9}{images/results/Lena512_midline7.png}{(b)}
%\subfig{0.3}{0.9}{images/results/Lena512_midline8.png}{(c)}
%\textbf{\\}
%\subfig{0.3}{0.9}{images/results/Lena512_midcol6.png}{(d)}
%\subfig{0.3}{0.9}{images/results/Lena512_midcol7.png}{(e)}
%\subfig{0.3}{0.9}{images/results/Lena512_midcol8.png}{(f)}
%\caption{
%Profiles of difference between the reconstruction and the true interior for the \textit{Lena} image.
%$x_0$, $\hat{x}$ and $x^\sharp$ are the padded FBP, the proposed reconstruction and the true interior, respectively.
%In blue: difference between the padded FBP and the true interior. In green: difference between the reconstruction with the proposed method with $\sigma = s = 3$ and the true interior.
%First row: profiles of the middle line of the image for (a) $R = 35$, (b) $R = 15$, (c) $R = 50$.
%Second row: profiles of the middle column for (d) $R = 35$, (e) $R = 15$, (f) $R = 50$
%}
%\label{fig:lenadiff}
%\end{pics}

\begin{figure}
% --------------------
\begin{minipage}{0.49\textwidth}
\begin{center}
\includegraphics[width=0.9\textwidth]{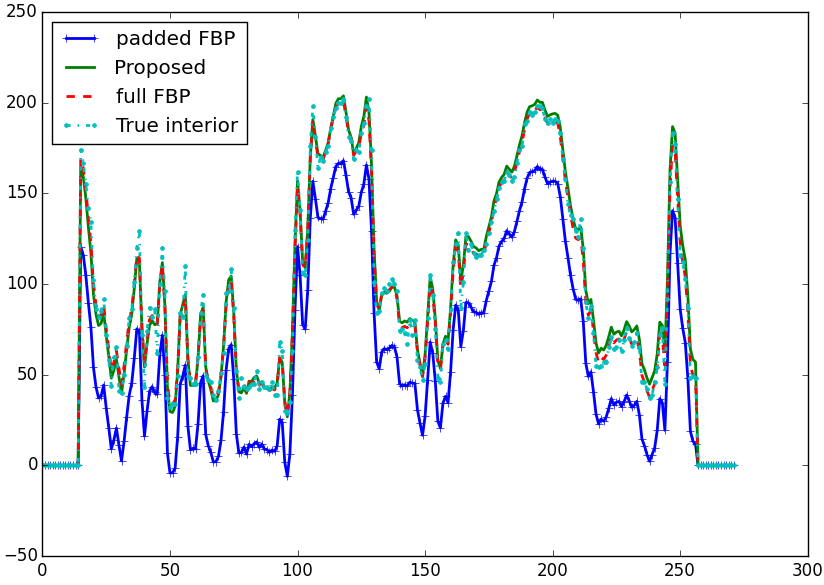}\\
(a)
\end{center}
\end{minipage}%
% --------------------
\begin{minipage}{0.49\textwidth}
\begin{center}
\includegraphics[width=0.9\textwidth]{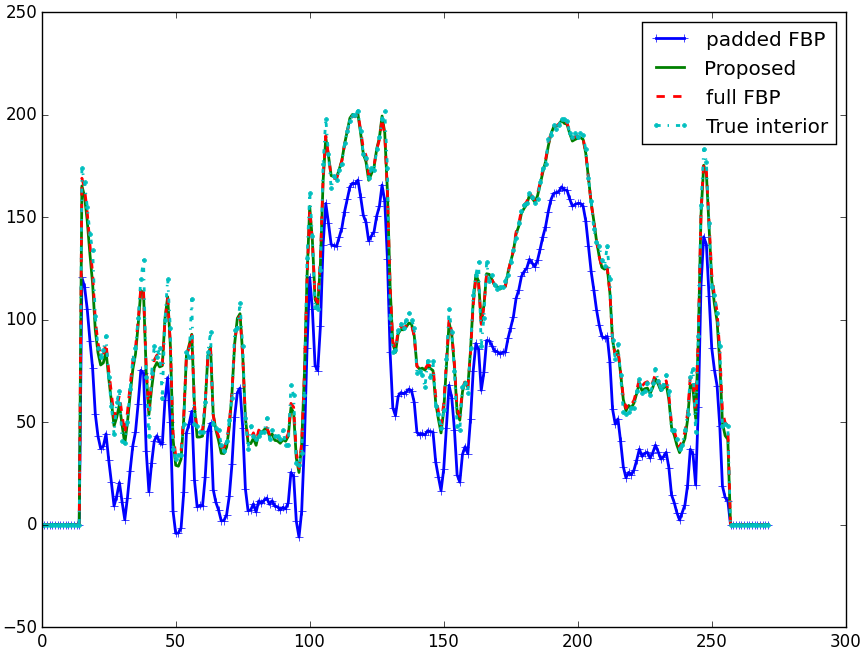}\\
(b)
\end{center}
\end{minipage}
% --------------------
\caption{
Line profiles of reconstructions with parameters $\sigma = s = 3$. (a) For radius = 15 pixels (b) For radius = 50 pixels.
}
\label{fig:lenaprofs}
\end{figure}

Figure \ref{fig:lenaprofs} highlights the quality improvement for higher values of $R$: the
reconstruction profile get closer to the true interior or the full FBP as $R$ increases.

% -----------------------------------------------------------------------------------------
% ----------------------------------------------------------------------------------------

%\begin{pics}
%\subfig{0.49}{0.9}{images/results/Lena512_midline7_prof_2.png}{(a)}
%\subfig{0.49}{0.9}{images/results/Lena512_midline8_prof_2.png}{(b)}
%\caption{
%Line profiles of reconstructions with parameters $\sigma = s = 3$. (a) For radius = 15 pixels (b) For radius = 50 pixels.
%}
%\label{fig:lenaprofs}
%\end{pics}

\begin{figure}
\begin{center}
\includegraphics[scale=0.4]{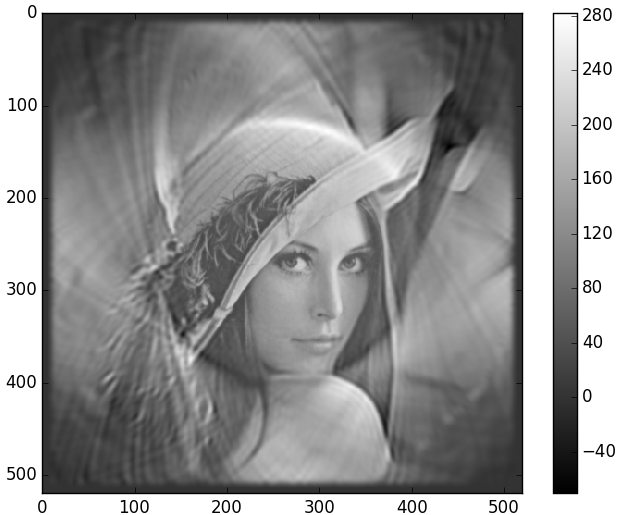}
\end{center}
\caption{
Extended image after solving \eqref{opt1}, without cropping to the region of interest.
The parameters used were $\sigma = s = 3$ and $R = 35$.
}
\label{fig:lenaext}
\end{figure}

It is also interesting to visualize the reconstruction of the whole extended image.
As it can be seen on Figure \ref{fig:lenaext}, the Gaussian basis even yields an approximation of the exterior.
This approximation is actually important for modeling the contribution of the external part in the acquired sinogram.
The bias correction is thus closely related to the modeling of the external part.

% -----------------------------------------------------------------------------------------
% -----------------------------------------------------------------------------------------

%\pic{0.4}{images/results/lena_ext_6.png}
%{
%\label{fig:lenaext}
%Extended image after solving \eqref{opt1}, without cropping to the region of interest.
%The parameters used were $\sigma = s = 3$ and $R = 35$.
%}

% -----------------------------------------------------------
% -------------------- Third test----------------------------
% -----------------------------------------------------------

\begin{figure}
% --------------------
\begin{minipage}{0.49\textwidth}
\begin{center}
\includegraphics[width=0.9\textwidth]{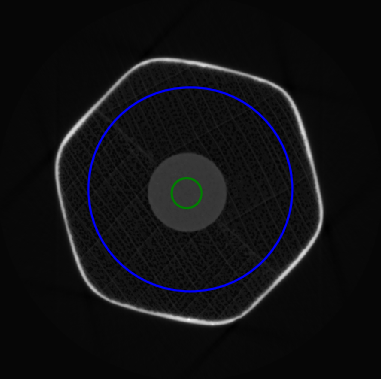}\\
(a)
\end{center}
\end{minipage}%
% --------------------
\begin{minipage}{0.49\textwidth}
\begin{center}
\includegraphics[width=0.9\textwidth]{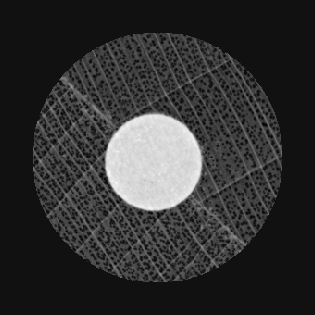}\\
(b)
\end{center}
\end{minipage}
% --------------------
\caption{
(a) \textit{Pencil} test image. In red: region of interest. In green: known sub-region.
(b) View of the region of interest.
}
\label{fig:crayon}
\end{figure}

The third test involves the image of a pencil resulting from a scan at the ESRF ID19 beamline, $512 \times 512$ pixels, shown on Figure \ref{fig:crayon}.
The width of the extended image is $N_2 = 520$.

% -----------------------------------------------------------------------------------------
% ----------------------------------------------------------------------------------------

%\begin{pics}
%\subfig{0.49}{0.9}{images/results/crayon_config_R20.png}{(a)}
%\subfig{0.49}{0.9}{images/results/volint_crayon.png}{(b)}
%\caption{
%(a) \textit{Pencil} test image. In red: region of interest. In green: known sub-region.
%(b) View of the region of interest.
%}
%\label{fig:crayon}
%\end{pics}

\begin{figure}
% --------------------
\begin{minipage}{0.3\textwidth}
\begin{center}
\includegraphics[width=0.9\textwidth]{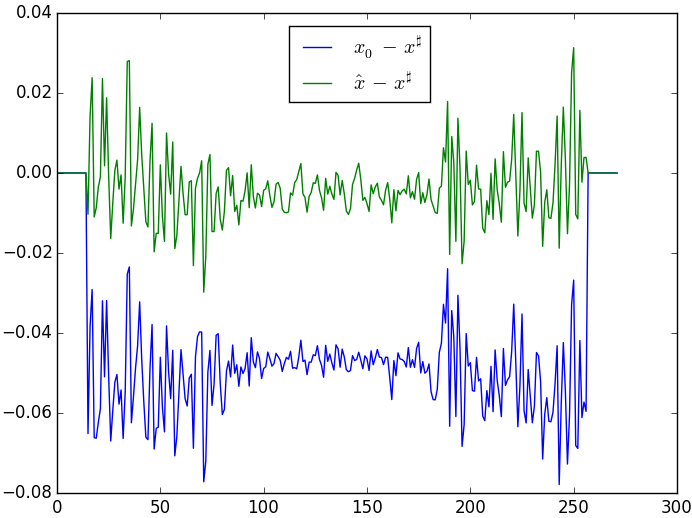}\\
(a)
\end{center}
\end{minipage}%
% --------------------
\begin{minipage}{0.3\textwidth}
\begin{center}
\includegraphics[width=0.9\textwidth]{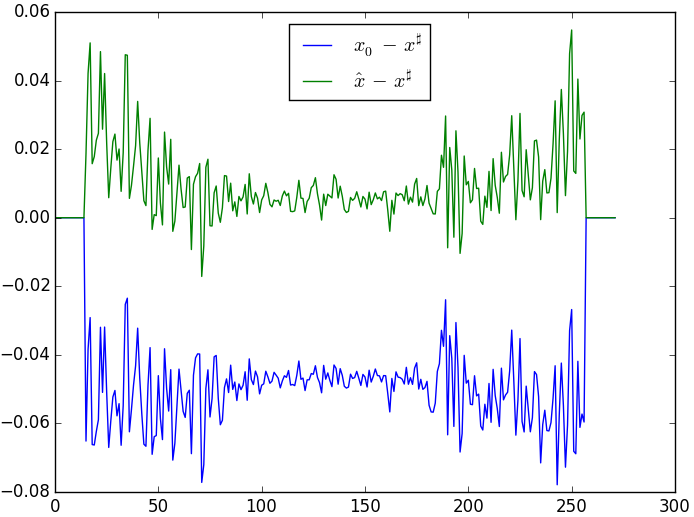}\\
(b)
\end{center}
\end{minipage}
% --------------------
\begin{minipage}{0.3\textwidth}
\begin{center}
\includegraphics[width=0.9\textwidth]{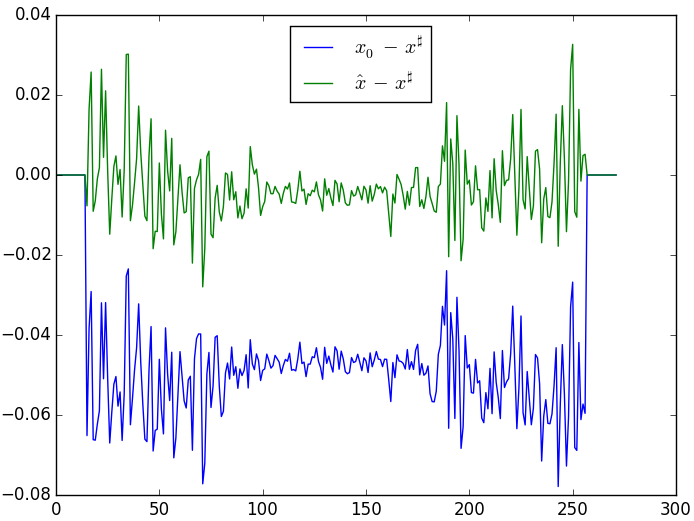}\\
(c)
\end{center}
\end{minipage}
\textbf{\\}
% --------------------
% --------------------
\begin{minipage}{0.3\textwidth}
\begin{center}
\includegraphics[width=0.9\textwidth]{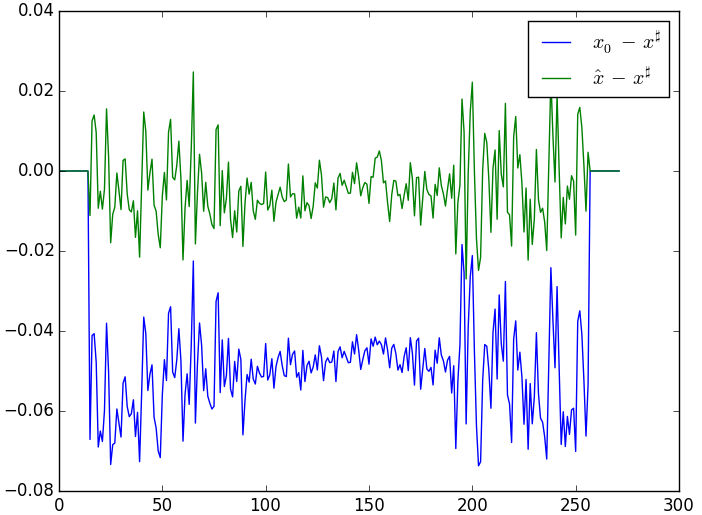}\\
(d)
\end{center}
\end{minipage}%
% --------------------
\begin{minipage}{0.3\textwidth}
\begin{center}
\includegraphics[width=0.9\textwidth]{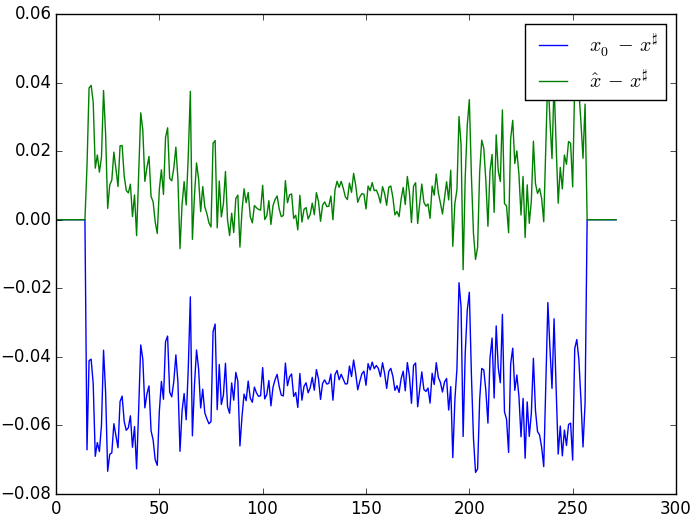}\\
(e)
\end{center}
\end{minipage}
% --------------------
\begin{minipage}{0.3\textwidth}
\begin{center}
\includegraphics[width=0.9\textwidth]{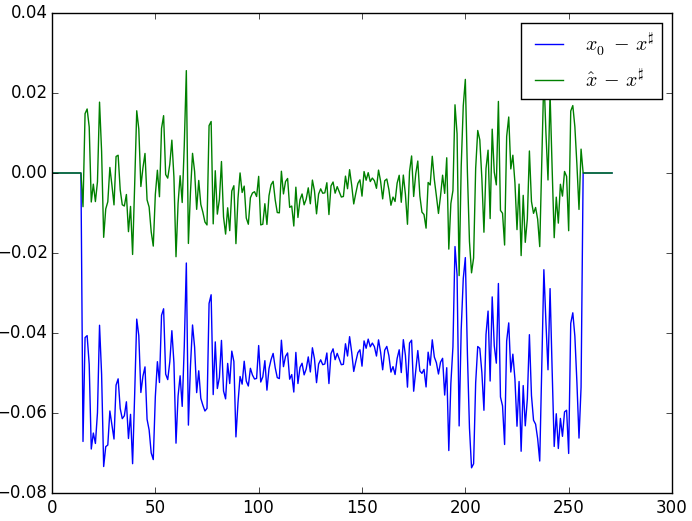}\\
(f)
\end{center}
\end{minipage}
% --------------------
\caption{
Profiles of difference between the reconstruction and the true interior for the \textit{pencil} image.
$x_0$, $\hat{x}$ and $x^\sharp$ are the padded FBP, the proposed reconstruction and the true interior, respectively.
In blue: difference between the padded FBP and the true interior. In green: difference between the reconstruction with the proposed method with $\sigma = s = 3$ and the true interior.
First row: profiles of the middle line of the image for (a) $R = 20$, (b) $R = 10$, (c) $R = 40$.
Second row: profiles of the middle column for (d) $R = 20$, (e) $R = 10$, (f) $R = 40$
}
\label{fig:crayondiff}
\end{figure}

\begin{figure}
\begin{center}
\includegraphics[scale=0.4]{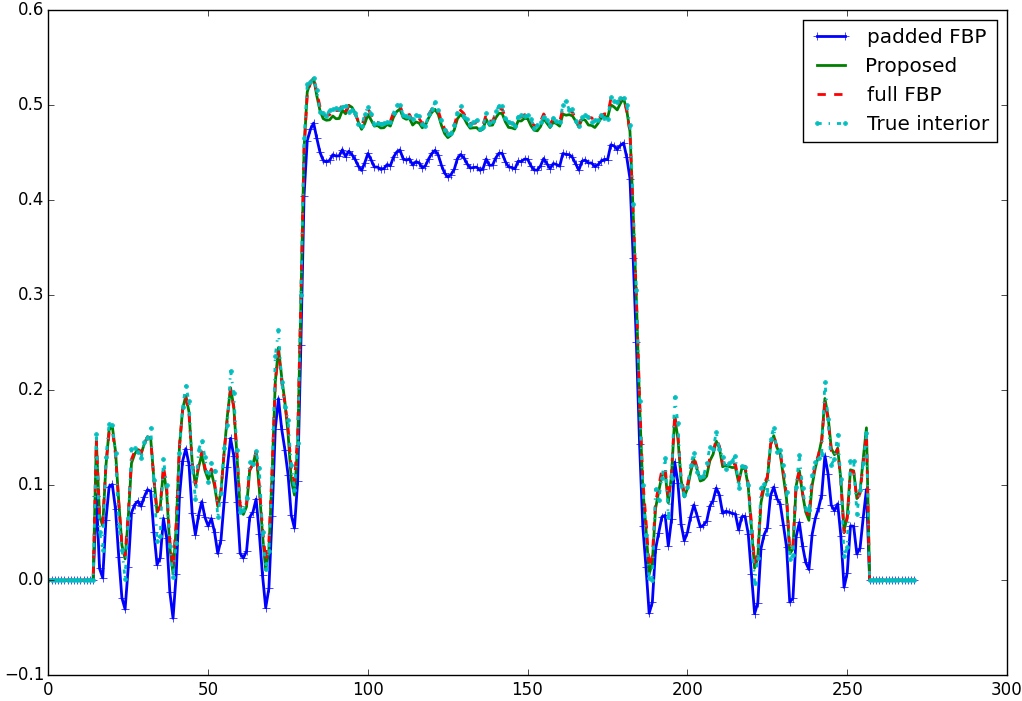}
\end{center}
\caption{
Line profiles for the \textit{pencil} image. The proposed method were applied with parameters $\sigma = s = 3$ and $R = 40$.
}
\label{fig:crayonprof}
\end{figure}

Figure \ref{fig:crayondiff}
shows profiles of the difference between the reconstruction and the true interior. 
On this image, a greater radius also improves the cupping removal. 
The profile of a line through the reconstructed image is depicted on Figure \ref{fig:crayonprof}.

% -----------------------------------------------------------------------------------------
% ----------------------------------------------------------------------------------------

%\begin{pics}
%\subfig{0.3}{0.9}{images/results/crayon3_midline.png}{(a)}
%\subfig{0.3}{0.9}{images/results/crayon4_midline.png}{(b)}
%\subfig{0.3}{0.9}{images/results/crayon5_midline.png}{(c)}
%\textbf{\\}
%\subfig{0.3}{0.9}{images/results/crayon3_midcol.png}{(d)}
%\subfig{0.3}{0.9}{images/results/crayon4_midcol.png}{(e)}
%\subfig{0.3}{0.9}{images/results/crayon5_midcol.png}{(f)}
%\caption
%{
%Profiles of difference between the reconstruction and the true interior for the \textit{pencil} image.
%$x_0$, $\hat{x}$ and $x^\sharp$ are the padded FBP, the proposed reconstruction and the true interior, respectively.
%In blue: difference between the padded FBP and the true interior. In green: difference between the reconstruction with the proposed method with $\sigma = s = 3$ and the true interior.
%First row: profiles of the middle line of the image for (a) $R = 20$, (b) $R = 10$, (c) $R = 40$.
%Second row: profiles of the middle column for (d) $R = 20$, (e) $R = 10$, (f) $R = 40$
%\label{fig:crayondiff}
%}
%\end{pics}

% -----------------------------------------------------------------------------------------
% -----------------------------------------------------------------------------------------

%\pic{0.4}{images/results/crayon5_midline_prof_2.png}
%{
%Line profiles for the \textit{pencil} image. The proposed method were applied with parameters $\sigma = s = 3$ and $R = 40$.
%\label{fig:crayonprof}
%}

\begin{figure}
% --------------------
\begin{minipage}{0.49\textwidth}
\begin{center}
\includegraphics[width=0.9\textwidth]{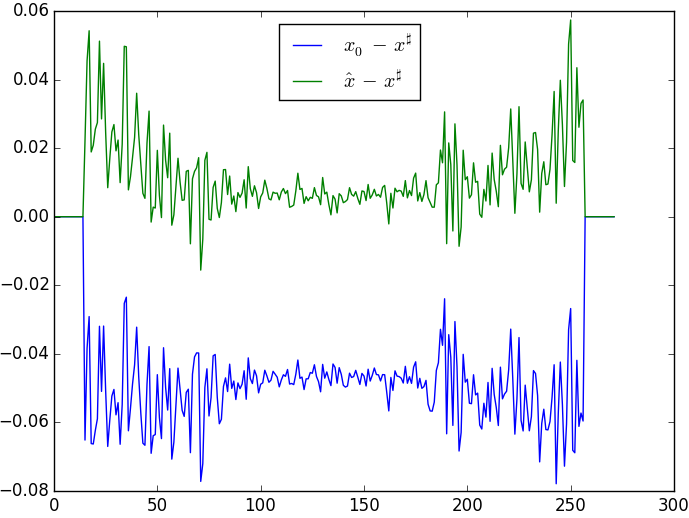}\\
(a)
\end{center}
\end{minipage}%
% --------------------
\begin{minipage}{0.49\textwidth}
\begin{center}
\includegraphics[width=0.9\textwidth]{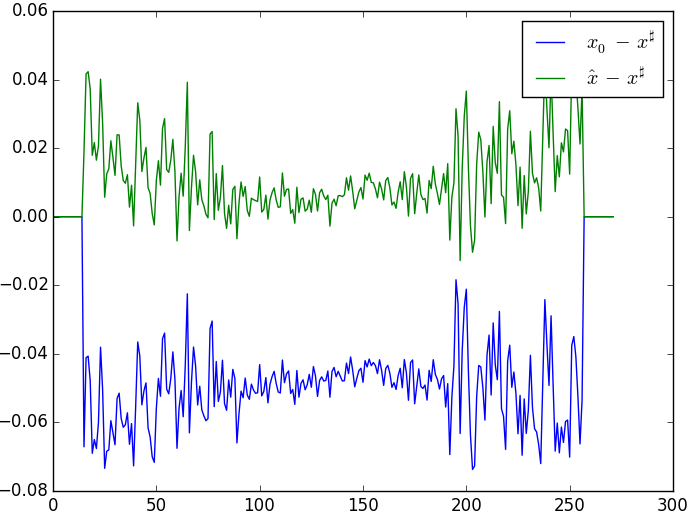}\\
(b)
\end{center}
\end{minipage}
% --------------------
\caption{
Profiles of difference between the reconstruction without known zone constraint and the true interior for the \textit{pencil} image.
(a) Line profile.
(b) Column profile.
}
\label{fig:crayonnoconstr}
\end{figure}

As a last remark, Figure \ref{fig:crayonnoconstr} shows the result of this method without using the known zone constraint, that is, 
without applying the constraint $g_{|\Omega_g} = g_0$ in \eqref{opt1}.
As expected, there is a not-null mean bias, even if it has been reduced with respect to padded FBP.

% -----------------------------------------------------------------------------------------
% ----------------------------------------------------------------------------------------

%\begin{pics}
%\subfig{0.49}{0.9}{images/results/crayon_noconst_line.png}{(a)}
%\subfig{0.49}{0.9}{images/results/crayon_noconst_col.png}{(b)}
%\caption
%{
%Profiles of difference between the reconstruction without known zone constraint and the true interior for the \textit{pencil} image.
%(a) Line profile.
%(b) Column profile.
%\label{fig:crayonnoconstr}
%}
%\end{pics}

Beside visual inspection, reconstructions can be quantitatively compared to the true interior of the test image.
Table \ref{table:psnr} shows the comparison with two
image metrics: peak signal to noise ratio (PSNR) and the structural similarity index (SSIM).
As these metrics are indicators of an \textit{average} distance between two images, we believe it is well suited for this
purpose of evaluating how the low frequencies are corrected by the proposed method.

\begin{table}
\begin{center}
\begin{tabular}{|c|c|c|c|c|}
\hline 
\textbf{Image} & \textbf{Reconstruction method} & \textbf{Parameters} & \textbf{PSNR} & \textbf{SSIM} \\ 
\hline 
Shepp-Logan & Padded FBP &  & 16.68 & 0.4578 \\ 
\hline 
Shepp-Logan & Proposed &  $\sigma = s = 3$, $R = 5$ & 26.74 & 0.6045 \\ 
\hline 
Shepp-Logan & Proposed & $\sigma = s = 3$, $R = 10$ & 26.56 & 0.6067 \\ 
\hline 
Lena & Padded FBP &  & 23.16 & 0.8605 \\ %23.51 & 0.8777 \\ 
\hline 
Lena & Proposed & $\sigma = s = 3$, $R = 15$ & 33.97 & 0.9560 \\ 
\hline 
Lena & Proposed & $\sigma = s = 3$, $R = 35$ & 35.47 & 0.9589 \\ 
\hline 
Pencil & Padded FBP &  & 26.41 & 0.8542 \\ 
\hline 
Pencil & Proposed & $\sigma = s = 3$, $R = 10$ & 31.15 & 0.9840 \\ 
\hline 
Pencil & Proposed & $\sigma = 4$, $s = 6$, $R = 40$ & 31.91 & 0.9901 \\
\hline
Pencil & Proposed & $\sigma = s = 3$, $R = 40$ & 34.22 & 0.9906 \\ 
\hline 
\end{tabular} 
\caption
{
Metrics of reconstruction quality for the three test images, computed inside the reconstructed ROI.
}
\label{table:psnr}
\end{center}
\end{table}

These results suggest that the proposed method yield better overall reconstruction quality than padded FBP.
In particular, it does not induce critical distortion in the reconstruction.

The following advantages of this method can be highlighted.
Using a Gaussian basis allows for efficient computation: the correction can be implemented as a convolution with a simple kernel.
This basis can also be extended to a multi-resolution grid, where big Gaussian functions are placed outside the ROI and
small Gaussian functions are placed inside the ROI, to reduce the degrees of freedom even further. 
This approach would be similar to the method proposed in \cite{local_wavelet_sparsity_besov}, 
although here only the correction is expressed in a multi resolution grid, not the image variables.
What can also be noted is that no assumption is done on the shape or location of the ROI and the known region: 
the known region can for example be several regions of various shapes, corresponding to pores in a sample.
Finally, by using the correction in the forward model \eqref{opt1}, the Helgason-Ludwig conditions are naturally fulfilled.

The proposed method depends on some parameters.
The first is the size of the extended image, which should be big enough to model the contribution of the external part.
% Ideally, this size should be close to the size of a detector that would cover the whole object.
The other parameters are the Gaussian standard deviation $\sigma$ and the spacing $s$ of the grid.
Both are related in a way that the Gaussian functions should slightly overlap to approximate constant functions: if $s$ value is high, then $\sigma$ should also be high and conversely. These parameters essentially tune how coarse is the Gaussian basis: high values 
would yield fast convergence but coarse result, while small values would yield slow convergence and fine result.

Using a Gaussian basis does not yield an exact correction of the error,
as Gaussian functions defined in Equation \eqref{g1d} do not form a basis.
For example, Gaussian functions do not yield a partition of unity, although a very close approximation of this property can be achieved \cite{gaussians_partition_unity}.
Thus, the final reconstruction cannot be \textit{exact} due to the basis coarseness, but
can provide results quite close to FBP with full data as seeing Figures \ref{fig:line1}, \ref{fig:lenaprofs} and \ref{fig:crayonprof}.

The fact that only a known zone of the ROI is enough to guarantee an almost-exact reconstruction might be counterintuitive,
especially in our case where this constraint is expressed in a coarse basis.
In the Gaussian basis, local constraints are propagated to the global image by the projection and backprojection operators involved in the process.
%propagate these constraints over the whole Gaussian grid: the solution of \eqref{opt1} should be consistent with the known zone constraint.
%
Using a coarse basis greatly reduces the degrees of freedom of problem \eqref{opt1}.
The classical tomographic reconstruction problem $P x = d$ turned into a least squares optimization $\argmin{x}{\norm{P x - d}_2^2}$
is ill-posed, even for complete data. In a local tomography setup, the ill-posedness is even worse \cite{local_tomoreconst21}.
Iterative solvers dealing with problem \eqref{opt2}
\begin{equation}\label{opt2}
\argmin{x}{\norm{C P x - d}_2^2 \quad \text{s.t.} \quad x_{|\Omega} = u_0}
\end{equation}
for $x$ in the pixel space, have very slow convergence in general due to the high number of degrees of freedom, even with spatial constraints.
The importance of reducing the degrees of freedom of \eqref{opt2} is highlighted for example in \cite{local_wavelet_sparsity_besov} and \cite{local_tv_gouillart}.

\section{Conclusion}
We presented a new technique of local tomography reconstruction based on the knowledge of a zone of the region of interest.
This technique corrects the cupping effect in an initial reconstruction by expressing the error in a coarse basis of Gaussian functions.
In accordance to local tomography uniqueness theorems, this method yield almost exact reconstructions, in spite of being only a correction with a coarse basis.
Besides, practical considerations are given for an efficient implementation suitable for reconstruction of real data.
A commented implementation of this method can be found at \cite{gitlocaltomo}.

     %-------------------------------------------------------------------------
     % The back matter of the paper - acknowledgements and references
     %-------------------------------------------------------------------------

     % Acknowledgements come after the appendices

%\ack{Acknowledgements}

     %-------------------------------------------------------------------------
     % TABLES AND FIGURES SHOULD BE INSERTED AFTER THE MAIN BODY OF THE TEXT
     %-------------------------------------------------------------------------

     % Simple tables should use the tabular environment according to this
     % model

%\begin{table}
%\caption{Caption to table}
%\begin{tabular}{llcr}      % Alignment for each cell: l=left, c=center, r=right
% HEADING    & FOR        & EACH       & COLUMN     \\
%\hline
% entry      & entry      & entry      & entry      \\
% entry      & entry      & entry      & entry      \\
% entry      & entry      & entry      & entry      \\
%\end{tabular}
%\end{table}

     % Postscript figures can be included with multiple figure blocks

%\begin{figure}
%\caption{Caption describing figure.}
%\includegraphics{fig1.ps}
%\end{figure}

     % References are at the end of the document, between \begin{references}
     % and \end{references} tags. Each reference is in a \reference entry.

%\begin{references}
%\reference{Author, A. \& Author, B. (1984). \emph{Journal} \textbf{Vol}, first page--last page.}
%\end{references}
\bibliographystyle{iucr} %plain, alpha, abbrv
\bibliography{biblio}

\end{document}